\documentclass[twocolumn,showpacs,preprintnumbers,amsmath,amssymb]{revtex4}

\usepackage{graphicx}
\usepackage{dcolumn}
\usepackage{bm}

\begin{document}

\title{Scale-free networks without growth}
\author{Yan-Bo Xie}
\author{Tao Zhou}
\author{Bing-Hong Wang}
\email{bhwang@ustc.edu.cn, Fax:+86-551-3603574.}
\affiliation{%
Department of Modern Physics and Nonlinear Science Center and ,
University of Science and Technology of China, Hefei, 230026, PR
China
}%

\date{\today}

\begin{abstract}
In this letter, we proposed an ungrowing scale-free network model,
wherein the total number of nodes is fixed and the evolution of
network structure is driven by a rewiring process only. In spite
of the idiographic form of $G$, by using a two-order master
equation, we obtain the analytic solution of degree distribution
in stable state of the network evolution under the condition that
the selection probability $G$ in rewiring process only depends on
nodes' degrees. A particular kind of the present networks with $G$
linearly correlated with degree is studied in detail. The analysis
and simulations show that the degree distributions of these
networks can varying from the Possion form to the power-law form
with the decrease of a free parameter $\alpha$, indicating the
growth may not be a necessary condition of the self-organizaton of
a network in a scale-free structure.
\end{abstract}


\maketitle

Many social, biological, and communication systems can be properly
described as complex networks with nodes representing individuals
and edges mimicking the interactions among them
\cite{Review1,Review2,Review3}. Examples are numerous: these
include the Internet, the World Wide Web, social networks,
metabolic networks, brain networks, food webs, and many others.
Recent empirical studies indicate that the networks in various
fields have some common characteristics, the most important of
which are called small-world effect \cite{WS} and scale-free
property \cite{BA}. The ubiquity of complex networks and the
discoveries of the common network properties inspire scientists to
construct general models. In the simplest way, these models can be
categorized into two classes, the growing models and the ungrowing
ones. In the growing models, the number of nodes in network grows
continuously, while in ungrowing cases, the number of nodes is
fixed. The Watts-Strogatz (WS) model is the pioneer and
representative one of ungrowing models, which can be constructed
by starting with a regular network and randomly moving one
endpoint of each edge with probability $p$ \cite{WS}. WS networks
display small-world effect, that is, they are highly clustered and
of small average distance. The most successful one of growing
models is the Barab\'{a}si-Albert (BA) model \cite{BA}, which
suggests that two main ingredients of the self-organization of a
network in a scale-free structure are growth and preferential
attachment. These points to the facts that most networks grow
continuously by adding new nodes, which are preferentially
attached to existing nodes with a large number of neighbors.

By using the mean-field theory on a toy model, Barab\'{a}si
\emph{et al} assert that the growth is one of the necessary
conditions for the emergence of scale-free property \cite{BAJ},
and the networks without growth may display an exponential degree
distribution. This hypothesis is now widely accepted. However,
some real-life networks having fixed size (or almost fixed size)
are of scale-free property. For instance, consider the friendship
networks of school children wherein the students are represented
by nodes and two nodes are connected by an edge if the
corresponding students have good friendship
\cite{friend1,friend2}. These networks do not grow with time since
the number of students in a class is almost fixed. Some students
have many good friends, while many others have only a few friends.
The students' heterogeneous societal ability leads to a skewed
degree distribution (or a skewed strength distribution for
weighted case \cite{friend2}), approximated to a power-law form. A
typical example in biological systems is the functional networks
of human brain \cite{Brain}. Although its structure varies with
time, the brain functional network's sizes is unchanged \cite{ex1}
while it displays scale-free property. Another significant
examples are the relationship networks of countries like world
trade web \cite{WTW1,WTW2} and world exchange arrangements web
\cite{WEAW1,WEAW2}. These networks are scale-free, while their
sizes almost do not change in the recent a few years.

Although these networks mentioned above have (almost) fixed size,
they are not static since their structures vary with time. These
motions can be considered as rewiring processes, that is, some
existing edges are removed while some new ones are added. For
example, in friendship networks, good friends may quarrel about
beliefs, money, or some other things, and become impassive to each
other; while some ones will become good friends for their common
interests and difficulties. The economic globalization promotes
different countries' economies to be integrated together in terms
of trade and capital flow, making countries more interdependent
than ever. In order to avoid financial crises, more flexible
economic regime should be implemented, and the government may
change the exchange-rate/trade regime for certain economic and
political reasons, leading redistribution of degree/weight in
world exchange arrangement networks and world trade networks.
These redistribution/rewiring processes make the structures
varying ceaselessly in recent years, but the scale-free property
is always observed. Notice that, in the ungrowing model proposed
by Barab\'{a}si \emph{et al} \cite{BAJ}, no edges will be removed
thus no rewiring processes will occur. In this letter, for the
first time to our knowledge, a scale-free network model without
growth is proposed, and we argue that the ungrowing networks can
have power-law degree distributions, which attributes to the
rewiring processes.

\begin{figure}
\scalebox{0.8}[0.9]{\includegraphics{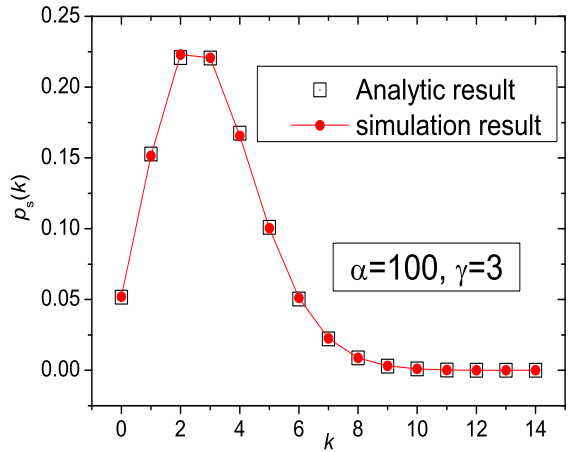}} \caption{(Color
online) The degree distribution of the present model for the case
$\alpha \gg 1$. The black hollow squares and red solid circles
represent the analytic and simulation results, respectively. The
analytic result agrees with the simulation accurately, and both of
them obey Possion forms. }
\end{figure}

For simplicity, we consider a network model with both fixed number
of nodes $N$ and fixed number of edges $E$. The initial state of
the network can be a random graph or any other types of graph.
Then the network evolves based on the following rewiring
processes: At each time step, an edge is randomly selected and
removed from the network.  At the same time, a node is selected
with the preferential probability $G(k_i)$, where $k_i$ is the
degree of the $i$th node. Another node is selected also with the
above preferential probability. Then a new edge connecting these
two nodes is created. Notice that in this rewiring process, the
total number of edges is unchanged. The above process is repeated
at each time step. Finally, we expect that the network reaches an
equilibrium state which is independent of the initial state.

Let $P(k,t)$ represents the number of nodes with degree $k$ at the
time step $t$.  $P(k,t)$ satisfies the following normalization
conditions
\begin{equation}
\sum_k P(k,t)=N,
\end{equation}
\begin{equation}
\sum_k kP(k,t)=2E,
\end{equation}
\begin{equation}
\sum_k G(k)P(k,t)=1.
\end{equation}
It is straightforward to write
down the master equation for $P(k,t)$,
\begin{eqnarray}
&&P(k,t+1)-P(k,t)={(k+1)P(k+1,t)-k P(k,t)\over E}\nonumber\\
&&+2G(k-1)P(k-1,t)-2G(k)P(k,t).
\nonumber\\
\end{eqnarray}
It may be helpful to explain the physical meaning of various terms
in the above equation.  The first two terms on the right hand side
represent the net gain of $P(k,t)$ due to the edge removing
process.  In particular, the first term represents the net gain of
$P(k,t)$ when the removing edge connects a node with the degree
$k+1$.  Notice that when an edge is randomly selected, the
connecting nodes have more chance to be of large degree.
Explicitly, the node with degree $k$ is selected with the
probability $k/\sum_k kP(k,t)=k/2E$.  Since each edge connects to
two nodes, the first term is thus $(k+1)P(k+1,t)/E$.  Similarly,
the second term represents the net loss of $P(k,t)$ when the
removing edge connects a node with the degree $k$. The third and
fourth terms represents the net gain of $P(k,t)$ due to the edge
adding process.  The third term represents the net gain of
$P(k,t)$ when the adding edge connects a node with the degree
$k-1$.  Notice that a node with degree $k$ is selected with the
probability $G(k)$. The fourth term represents the net loss of
$P(k,t)$ when the adding edge connects a node with the degree $k$.

When $t$ is sufficiently large, we expect that $P(k,t)$ approaches a
stationary state denoted by $P_s(k)$ that satisfies
\begin{widetext}
\begin{equation}
{(k+1)P_s(k+1)-kP_s(k)\over 2E}+G(k-1)P_s(k-1)-G(k)P_s(k)=0,
\end{equation}
\end{widetext}
where $P_s(k)$ also satisfy the normalization conditions
Eqs.(1-3).  Notice that this master equation is a second order
equation in which $P_s(k+1)$ is determined by $P_s(k)$ and
$P_s(k-1)$. Eqs.(1-3,5) can be solved by the following method.
Define
\begin{equation} H(k)=G(k)P_s(k)
\end{equation} and
\begin{equation} W(k)=kP_s(k)/2E.
\end{equation}
Then Eqs.(2-3,5) can be rewritten as
\begin{equation}
\sum_k W(k)=1, \end{equation} \begin{equation} \sum_k
H(k)=1,\end{equation} \begin{equation} H(k)=H(k-1)+W(k+1)-W(k).
\end{equation}
By iteration, one can obtain \begin{equation}
H(k)=H(0)+W(k+1)-W(1).
\end{equation}
Since $W(0)=0$ and $$\sum_k H(k)=\sum_k W(k)=1,$$ one immediately
have \begin{equation} W(k+1)=H(k) \end{equation} for $k\geq 0$.
Substituting Eqs.(6-7) into Eq.(12), one obtains the following
solution
\begin{equation}
P_s(k)={(2E)^k\over k!}G(k-1)G(k-2)...G(0)P_s(0)
\end{equation}
for $k\geq 1$ with $P_s(0)$ determined by the normalization
condition Eq.(1).  It should be stressed that given the dependence
of $G(k)$ on $k$, there is still a proportional coefficient in
$G(k)$ determined by Eq.(3).

Hereinafter, we focus on a special form, also one of the simplest
cases, of $G(k)$, where $G(k)$ is linearly correlated with $k$,
\begin{equation}
G(k)=\frac{k+\alpha}{\sum^N_{i=1}(k_i+\alpha)}={k+\alpha\over
2E+N\alpha},
\end{equation}
where $k_i$ denotes the degree of node $i$, and $\alpha$ is a
constant (Notice that $G(k)$ satisfies the normalization condition
Eq. (3).). Substituting the above equation into Eq. (13), one can
straightforwardly obtain, in this special case, that
\begin{equation}
P_s(0)=N(\frac{\alpha}{\alpha+\gamma})^\alpha
\end{equation}
\begin{equation}
P_s(k)=N({\alpha\over \alpha+\gamma})^{\alpha} ({\gamma\over
\alpha+\gamma})^k{1\over k!}\alpha(\alpha+1)...(\alpha+k-1), k\geq
1
\end{equation}
where $\gamma=2E/N$.

When $\alpha \gg 1$, the mechanism of preferential attachment is
destroyed, that is, since $\alpha$ probably plays the main part in
the sum $\alpha+k_i$, the probability a high degree node being
selected is approximately the same as that of a low degree node
(see Eq. 14). Therefore, one expects that the present network will
approach a completely random network with its degree distribution
obeying the Possion distribution \cite{RG}
\begin{equation}
P_s(k)=N\frac{\texttt{e}^{-\gamma}}{k!}\gamma^k.
\end{equation}
When $\alpha \ll 1$ and $\gamma \ll 1$, most nodes are isolated
and the nodes with large degree are very few. However, when
$\alpha \ll 1$ and $\gamma \succeq 1$, interesting result emerges.
The corresponding degree distribution obeys a power-law form with
exponent approximated to 1 when $k\ll \gamma / \alpha$
\begin{equation}
P_s(k)=N(\frac{\alpha}{\gamma})^\alpha\frac{\alpha}{k}.
\end{equation}

\begin{figure}
\scalebox{0.8}[0.9]{\includegraphics{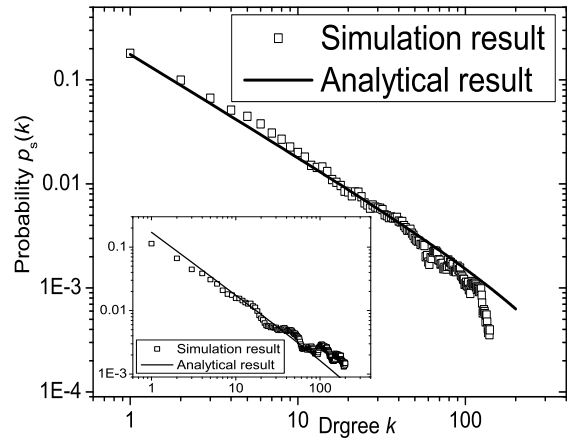}} \caption{The degree
distribution of the present model with $\alpha=0.01$ and
$\gamma=5$. The hollow squares and solid curve represent the
simulation and analytical results, respectively. The analytical
result agrees with the simulation well, obeying an approximately
power-law form. The inset exhibits the case of $\alpha=0.01$ and
$\gamma=10$ for comparison.}
\end{figure}

In succession, we will show some simulations. $10^6$ time steps
are performed in all simulations while only the final $2\times
10^5$ time steps are used for statistical average. The initial
graph is simply the completely random graph with given $N$ and
$E$. The network size if fixed as $N=1000$ unless a special
statement is addressed. In figure 1, we report the case for very
large $\alpha$, where $p_s(k):=P_s(k)/N$ denotes the normalized
probability function for degree distribution. The simulation
result strongly supports the analytic one (see Eq. (16)), and both
of them obey Possion forms. In this figure, one will find that
there are about 5\% nodes with degree zero. However, this
probability only reflects the average number of isolated nodes
over all configurations. Because of the rewiring mechanism, no
nodes are always isolated. Although real-life networks are not
always connected, most previous studies have focused on connected
graphs. Therefore, hereinafter, we neglect the isolated nodes.

\begin{figure}
\scalebox{0.8}[0.9]{\includegraphics{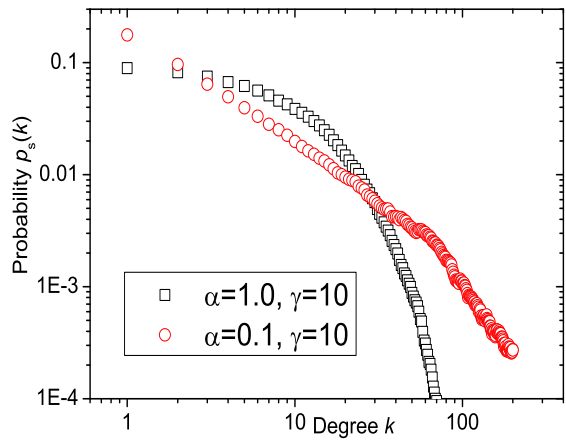}} \caption{(Color
online) The degree distributions of the present model. The
parameter $\gamma=10$ is fixed, and the black squares and red
circles represent the cases with $\alpha=1.0$ and $\alpha=0.1$,
respectively. The degree distribution with $\alpha=1.0$ obeys
approximately an exponential form rather than a power-law form.}
\end{figure}

In figure 2, we report the analytical and simulation results about
degree distributions generated by very small $\alpha=0.01$ and
$\gamma=5$. The degree distribution displays obviously scale-free
property with a cut-off at $k \approx 120$. The solid curve
denotes the analytical solution, which agrees with the simulation
well. We also have checked that the departure from Eq.(18) will
get greater if the parameter $\alpha$ becomes larger. In the
inset, the case of $\gamma=10$ is shown for comparison. As
mentioned above, the two extreme cases with $\alpha\approx 0$ and
$\alpha\gg 1$ will lead to the power-law and Possion
distributions, respectively. In figure 3, we report the simulation
results for $\alpha=0.1$ and $\alpha=1.0$. The departure from the
power law becomes larger as the increase of $\alpha$, and the
degree distribution with $\alpha=1.0$ obeys approximately an
exponential form rather than a power-law form.

In conclusion, we proposed an ungrowing model and obtained the
analytic results for arbitrary forms of selection probability
$G(k)$. One special kind of the present networks is investigated
in detail, which can generate networks with different classes of
degree distributions by tuning two parameters. Especially, when
$\alpha$ is close to zero and $\gamma \succeq 1$, the present
networks display scale-free property with its exponent
approximately equal to that of world exchange arrangement web
\cite{WEAW2}.

The previous empirical studies have demonstrated the extensive
existence of scale-free networks. Most of them are growing at all
times, while some others are of (almost) fixed size. Most previous
models \cite{Review1,Review2,Review3,BA,BAJ} suggested the growth
a key ingredients of the emergence of scale-free structure, thus
fail to provide an underlying mechanism by which the ungrowing
networks exhibiting scale-free property. Here we argue that the
rewiring mechanism widely existed in real-life networks like
friendship networks \cite{friend1,friend2}, brain functional
networks \cite{Brain}, world trade web \cite{WTW1,WTW2}, and world
exchange arrangements web \cite{WEAW1,WEAW2}, may be the origin of
the emergence of scale-free structure in fixed-size networks.

Finally, we would like to point out that many different classes of
degree distributions can be obtained from different forms of
$G(k)$ \cite{unpub}. In particular, one can obtain the power law
degree distribution $P_s(k)\sim k^{-\beta-1}$ with tunable
exponent $\beta$ by setting $G(k)\sim
\frac{k^{\beta+1}}{(k+1)^\beta}$ when the number of edges $E$ is
properly chosen \cite{unpub}.

This work is support by the National Natural Science Foundation of
China under Nos. 70271070, 10472116, 70471033, and 70571074, and
the Specialized Research Fund for the Doctoral Program of Higher
Education under No. 20020358009.


\begin{thebibliography}{Reviews}
\bibitem{ex1} In human brain functional networks, each node
represents a voxels of dimension $3\times 3.475\times 3.475
\texttt{mm}^3$ in the picture taken by functional magnetic
resonance imaging (FMRI) in humans. Therefore, the network size is
fixed as $N=36\times 64\times 64$.
\end{thebibliography}
\end{document}